\documentclass{aa}  

\usepackage{graphicx}
\usepackage{natbib}
\bibpunct{(}{)}{;}{a}{}{,} % to follow the A&A style

\begin{document}

\title{Ray-tracing 3D dust radiative transfer with \textsc{DART-Ray}: code upgrade and public release}

\author{Giovanni Natale\inst{1} \and Cristina C. Popescu\inst{1,2,3} \and Richard. J. Tuffs\inst{3} \and Adam J. Clarke\inst{1} \and Victor P. Debattista\inst{1} \and J\"org Fischera\inst{3} \and Stefano Pasetto\inst{4}\and Mark Rushton\inst{2} \and Jordan J. Thirlwall\inst{1}}

\institute{University of Central Lancashire, Jeremiah Horrocks Institute, Preston, PR1 2HE, UK \\
\email{gnatale@uclan.ac.uk} \and 
The Astronomical Institute of the Romanian Academy, Str. Cutitul de Argint 5, Bucharest, Romania \and  
Max Planck Institute f\"{u}r Kernphysik, Saupfercheckweg 1, D-69117 Heidelberg, Germany \and
The Carnegie Observatories - Carnegie Institution for Science, 813 Santa Barbara St, Pasadena, CA 91101, USA }

\abstract
{We present an extensively updated version of the purely ray-tracing 3D dust radiation transfer code DART-Ray. The new version includes five major upgrades : 1) a series of optimizations for the ray-angular density and the scattered radiation source function; 2) the implementation of several data and task parallelizations using hybrid MPI+OpenMP schemes; 3) the inclusion of dust self-heating; 4) the ability to produce surface brightness maps for observers within the models in HEALPix format; 5) the possibility to set the expected numerical accuracy already at the start of the calculation. We tested the updated code with benchmark models where the dust self-heating is not negligible. Furthermore, we performed a study of the extent of the source influence volumes, using galaxy models, which are critical in determining the efficiency of the DART-Ray algorithm. The new code is publicly available, documented for both users and developers, and accompanied by several programmes to create input grids for different model geometries and to import the results of N-body and SPH simulations. These programmes can be easily adapted to different input geometries, and for different dust models or stellar emission libraries.}

\titlerunning{DART-Ray upgrade and public release}

% \keywords{giant planet formation --
%                $\kappa$-mechanism --
%                stability of gas spheres
%               }
%\end{abstract}

\maketitle

\section{Introduction}

The modelling of observations of astrophysical objects in the wavelength range from the UV to the submm is a challenging task. For a vast variety of scales, from proto-planetary systems to galaxies, the emission in this wavelength range is dominated either by primary sources of radiation (e.g. stars or active galactic nuclei, AGN), predominant in the UV and optical, or by the re-emission of absorbed photons by interstellar dust, predominant in the mid- and far-infrared. Usually, the near-infrared range is a region of transition between the two kinds of emissions.

The observed emission produced by the primary sources and the dust is mutually affected. On the one hand, the dust dims and scatters the light generated by the stars or AGNs. On the other hand, the light from the primary sources heats the dust, determining its temperature and thus its emission spectra. Furthermore, although most astrophysical objects are optically thin at long infrared wavelengths, the dust emission produced at one location can also be absorbed and scattered by the dust located elsewhere, a process often referred to as dust `self-heating'. 

Performing dust radiative transfer (RT) calculations is the essential step to reproduce the observations in a (as much as possible) self-consistent way. The problem is computationally challenging because of its non-locality (in the spatial, angular direction, and wavelength dimensions) and non-linearity (e.g. the dust emission spectra depends non linearly on the absorbed luminosity). Furthermore, the presence of six independent variables (three spatial coordinates, two angular coordinates and the wavelength) makes it very challenging to handle the large memory required if the 3D dust radiation transfer equation has to be solved directly (see \citet{Steinacker13} for a recent review). 

For all the above reasons, the vast majority of the dust radiative transfer codes adopt a Monte-Carlo (MC) approach (e.g. see lists of codes in \citet{Steinacker13, Pascucci04, Pinte09} and \citet{Gordon17}, hereafter G17), which is an elegant, flexible way to follow the propagation of light within dusty objects and it heavily reduces the memory requirements since there is no need to store the intensity as a function of the angular direction at each spatial position. The basic Monte-Carlo approach is not efficient in producing images at specific line-of-sights but this problem can be solved by combining it with a ray-tracing procedure called peel-off \citep{Yusef-Zadeh84}. Other acceleration techniques to maximize the use of photon particles are available \citep{Steinacker13}. 

Despite all the known technical difficulties, in the last few years we have been developing a 3D dust RT code that is purely ray-tracing, that is, it does not make any use of MC techniques but is simply based on the calculation of the radiation intensity variation along numerous directions chosen deterministically by the code algorithm. The code, named DART-Ray, and its basic algorithm have been introduced in \citet[][hereafter NA14]{Natale14}. As stated in that article, the main motivation to develop this code is having a specific tool for the calculation of the radiation field energy density (RFED) throughout any region of the RT model under consideration. To this goal, our group already made extensive use of a ray-tracing RT code based on the scattering intensity approximation of \citet{Kylafis87} \citep[see e.g.][]{Popescu2000, Popescu11}. However, this code can be applied only to axisymmetric galaxy models, while DART-Ray is able to handle any geometry.

RFED are obviously also calculated by MC codes, but the main focus of MC codes is the production of surface brightness images that might or might not require the RFED being accurately calculated at all positions and all wavelengths. In particular, our focus on the RFED is due to the importance of this quantity in other fields of astrophysics, such as high-energy astrophysics, where it is necessary to calculate the radiation due to inverse-Compton of cosmic rays interacting with the interstellar radiation field produced by stellar and dust emission.

Furthermore, the DART-Ray algorithm is not a brute-force ray-tracing algorithm. It takes advantage of a so far not well-studied property of the radiation sources within RT models, that is, that these sources often do not contribute significantly to the RFED everywhere but only within a fraction of it called the source influence volume. Although being essentially a cell-to-cell radiation transfer algorithm, the gain in efficiency of DART-Ray with respect to a brute force algorithm comes from its method to estimate the extent of the source influence volumes and perform radiation transfer calculations only within them. The extent of these volumes in astrophysical objects and the possible advantages that can be exploited in radiation transfer codes have never been clarified. Intuitively, in dusty objects the extent of this volume could be quite reduced relative to the size of the models, especially for the scattered light sources, which are low intensity sources compared to the sources actually producing radiation, such as stars and dust thermal emission. DART-Ray allows one to examine the extent of the source influence volumes and thus verify when they are small relative to the entire model size. Finally,  handling 3D dust radiative transfer in a manner different to those of the widely used MC techniques provides a useful test for the reliability of scientific results obtained by MC codes. In principle, an agreement between two or more MC codes could be due to the adoption of the same numerical method, but this interpretation can be discarded when a different RT solver obtains the same result. This kind of comparisons is already underway with the TRUST benchmark project in which DART-Ray is participating (see G17). 

The code presented in NA14 was a good first step in the development of a mature code, but there was some scope for improvement to ameliorate several limitations: firstly it could only be executed one wavelength at a time, neglecting dust self-heating, which requires multi-wavelength runs; secondly parallelization was implemented only for shared memory machines; thirdly the inaccuracy from the blocking of the rays could not be set at the start of the calculation and could only be measured by re-running the model with a different value for a threshold parameter ($f_U$ , see Na14 or below); fourthly stochastically heated dust emission was excluded from the calculation \citep[this was added in][]{Natale15}. Furthermore, a substantial reduction in the execution time could have been achieved through the implementation of a more efficient algorithm for the optimization of the ray angular density. 

In this paper, we present a new version of the code (hereafter DART-Ray V2), which is a substantial improvement to the one presented previously. The new code is publicly available and documented for both users and developers\footnote{https://github.com/gnatale/DART-Ray}. As well as addressing the issues highlighted above, we have added new features, such as the ability to create `internal observer' maps viewed from within the RT models in HEALPix format (which can be used for Milky Way studies) and at arbitrary lines of sight, without repeating the radiation transfer calculation. This latter feature can be used, for example, to create animations for the presentation of the results.

The structure of the paper is the following. In Sect. \ref{dartray_algo} we briefly summarize the radiation transfer algorithm used in DART-Ray. In Sect. \ref{update_descriptions} we describe the numerous updates to the code.  In Sect. \ref{benchmark_comparison} we show the comparison of the code with benchmark solutions including dust self-heating. In Sect. \ref{source_influence_vol} we present a study of the extent of the source influence volumes for radiation sources within different galaxy models. In Sect. \ref{Pro_and_cons} we discuss the advantages and disadvantages of the DART-Ray algorithm and in Sect. \ref{sec_improvements} its possible further improvements.

\section{The DART-Ray dust radiation transfer algorithm}
\label{dartray_algo}

The general strategy of the RT algorithm of DART-Ray V2 is the same as the one presented in NA14. However, there are many differences regarding the implementation and the newly added capabilities (see Sect. \ref{update_descriptions}). Here we make a brief summary of the RT algorithm, highlighting the main steps. We encourage users of the code and readers interested in more specific details to read the user  guide and the code documentation on the code webpage as well as sections 2 and 3 of NA14 for further clarifications on specific points. 

In DART-Ray, a RT model is subdivided into an adaptive 3D Cartesian grid of cells, each with a given input value of stellar volume emissivity $j_{\lambda}(\vec{r})$ (luminosity per unit volume per unit frequency and per unit solid angle) and dust optical depth per unit length $k_\lambda\rho_{\rm d}(\vec{r})$ (with $k_\lambda$ the extinction coefficient and $\rho_d$ the dust density). The albedo $\omega_\lambda$ and the anisotropy parameter of the Henyey-Greenstein phase function $g_\lambda$ are determined by the assumed dust model. Given these input quantities, the code calculates: 
\begin{itemize}
\item the RFED $U_\lambda (\vec{r})$ for each cell;
\item  the scattered luminosity source function $j_{\lambda, sca}(\vec{r},\theta, \phi)$ which contains the scattered radiation luminosity per unit volume and per unit solid angle for each dusty cell. In general, the scattered luminosity is not isotropically distributed and thus depends on the angular direction ($\theta, \phi$); 
\item the dust emission source function $j_{\rm \lambda,d}(\vec{r})$ which contains the luminosity per unit volume and per unit solid angle produced in each cell containing dust;
\item the specific intensity $I_{\lambda,obs}(\vec{r}, \theta, \phi)$ of the radiation produced by each cell/point source, and reaching the observer located either far away or within the RT model.  It is derived from the source terms $j_{\lambda}(\vec{r})$,   $j_{\lambda, sca}(\vec{r},\theta, \phi)$ and  $j_{\rm \lambda,d}(\vec{r})$, and the optical depth between the cell/point source and the observer.  It can be used to calculate surface brightness maps at the position of the observer.  

\end{itemize}

The code performs first the RT calculation for the stellar emission and subsequently that for the dust emission (the latter added in this code version, see Sect. \ref{self_heating}). In both cases, the RT algorithm is subdivided in three steps: \\
1) the determination of a lower limit $U_{\lambda, \rm LL}(\vec{r})$ to the RFED distribution $U_{\lambda}(\vec{r})$; \\
2) the processing of radiation coming directly from radiation sources; \\
3) the processing of radiation scattered by dust.

In all these steps, the DART-Ray algorithm considers one radiation source at a time (that is, either an `emitting cell' whose stellar or dust volume emissivity is not zero or a point source). For each source, it calculates the contributions of the radiation emitted by the source to the RFED within a certain volume surrounding it. In steps 2 and 3 the contributions to $j_{\lambda,sca}(\vec{r},\theta, \phi)$ for each cell of this volume are also calculated. The value of these contributions is derived after each ray-cell intersection (see Sect. 3.2 in NA14). In the new code version, the ray tracing from a radiation source within the surrounding volume involves a ray angular density optimization procedure described in Sect. \ref{ray_density_optimization}. 

During step 1, the volume considered around each radiation source has a fixed extent chosen by the user (typically 10-20\% of the entire model size). In this way, a lower limit of the RFED distribution $U_{\lambda, \rm LL}(\vec{r})$ is derived because the contributions in the regions beyond these volumes are not taken into account. 

In step 2, the ray-tracing calculation is performed once again from the beginning but this time the rays originating from the radiation sources are blocked if 
 the ray contribution $\delta U_\lambda$ to the local RFED is `negligible' at all wavelengths, that is, when 
\begin{equation}
\delta U_\lambda(\vec{r}) < f_{\rm U} U_{\lambda,\rm LL}(\vec{r}), 
\label{equ_blocking_criteria}
\end{equation} 
 where $f_{\rm U}$ is a threshold parameter chosen indirectly by the user depending on the desired numerical accuracy (see Sect. \ref{expected_num_accuracy}). 
 
Finally, during  step 3 the scattered radiation stored within the dusty cells is processed. In opposition to step 2, the radiation produced by emitting cells is typically direction-dependent, since the assumed scattering phase function (Henyey-Greenstein) is in general not isotropic\footnote{Scattering is particularly anisotropic in the UV and optical wavelength regimes, while it is almost isotropic in the infrared \citep{Draine03}.}. Nonetheless, apart from few technical differences, the calculation during this step proceeds essentially as in step 2\footnote{One important difference is that the value of $U_{\lambda, \rm LL}(x)$ is updated firstly with the RFED distribution found at the end of step 2 and then with that found at the end of each scattering iteration.}. Since scattered radiation can be scattered multiple times, several scattering iterations are needed. These iterations are stopped when the remaining scattered radiation luminosity waiting to be processed is only a small fraction $f_L$ of the total scattered luminosity of the model as found at the end of step 2. 

The code performs firstly the radiation transfer calculations only for the stellar emission. Then, it starts the calculation for the dust emission. The dust emission spectra produced by each dusty cell can be derived from the luminosity absorbed by the dust (which depends on $U_{\lambda}(\vec{r})$, the dust density and the dust opacity coefficients). The radiation emitted by dust undergoes the same type of propagation as for the stellar emission with the difference that the extra-radiation absorbed in this process affects the dust temperature and thus its emission. Therefore, since dust emission and absorption are coupled, multiple iterations of the entire radiation transfer procedure described in this section are performed until the dust emission spectra have converged at all positions (see Sect. \ref{self_heating}).   
 
Once $j_{\lambda}(\theta, \phi)$, $j_{\lambda, sca}(\theta, \phi)$ and $j_{\rm \lambda,d}$ are known, one can calculate the specific intensity $I_{\lambda,0}$ of the radiation departing from each source along any angular direction (see equation 7 of NA14). The specific brightness for the radiation arriving to the observer is then simply $I_{\lambda,obs} = I_{\lambda,0}e^{-\tau_\lambda}$ with $\tau_\lambda$ the optical depth between the source and the observer position. The code calculates $I_{\lambda,obs}$ for all cells and point sources and then use volume rendering techniques to produce surface brightness maps. We note that, if the source functions are saved, $I_{\lambda,obs}(\vec{r}, \theta, \phi)$ can be calculated for arbitrary observer positions without repeating the entire RT calculation.      

\section{Update descriptions}
\label{update_descriptions}

\subsection{Optimizations}

Compared to NA14, we implemented two main changes affecting the code speed and memory requirement. These are an improved algorithm for the optimization of the ray angular density and the implementation of wavelength-dependent angular resolution for the scattering source function $j_{\rm \lambda,sca}$. 

\subsubsection{Ray angular density optimization}
\label{ray_density_optimization}

DART-Ray performs ray-tracing operations from each radiation source (either an emitting cell or a point source) throughout a 3D Cartesian adaptive grid. When a source contributes significantly to the RFED within a grid cell, it is necessary that at least several rays, originating from the source, intersect the cell in order to achieve good numerical accuracy for the RFED and the source functions $j_{\rm \lambda, sca}$ and $j_{\rm \lambda,d}$ at that grid position. In this way, one also avoids missing cells at similar distances. Unfortunately, the extent of the source influence volume cannot be known in advance. Therefore, it is not possible to set a sufficiently high ray angular density (the number of rays per unit solid angle) right at the beginning of the ray propagation, so that all the cells within the source influence volume are properly intersected. Instead, the ray angular density has to be derived iteratively while the rays propagate through the model. The optimal ray angular density is also not necessarily uniform over the entire solid angle, as seen from the radiation source, but it can well be direction--dependent. Furthermore, the variable cell size of the adaptive 3D grid of emissivity and opacity can make the optimal angular ray density vary with the distance from the radiation source. 

The directions along which the rays are cast are those defined by the lines passing through the source position and the centres of the spherical pixels of a concentric sphere, subdivided according to the HEALPix sphere pixelation scheme \citep{Gorski05}. The advantage of using HEALPix is that the angular resolution of the sphere pixelation (defined as a quad-tree) can be varied easily and there are fast routines available for spherical pixel searches (e.g. to obtain spherical angles from the pixel number and vice versa). The basic idea is to start the RT calculation by using a low initial HEALPix resolution. While a ray is propagating throughout the model, the code can vary the ray angular density by moving from one HEALPix resolution level to the immediately higher or lower as described below. Because of the quad--tree structure of HEALPix, any change in HEALPix resolution corresponds to a factor 4 in the variation of the ray angular density.  

Specifically, the ray angular density has to be increased when the following two conditions are met: 

\begin{enumerate}

\item the ray beam size is larger than the maximum allowed size, that is:

\begin{equation}
\Omega_{\rm HP, EM} > \frac{\Omega_{\rm INT}}{\rm N_{\rm rays}},
\label{equ_omega_high}
\end{equation}

where $\Omega_{\rm HP, EM}$ is the solid angle of the beam associated with a ray, $\Omega_{\rm INT}$ is the solid angle subtended by the last intersected cell  and N$_{\rm rays }$ is the minimum number of rays that has to intersect a cell (input-defined); 
\item the ray has either not reached the boundary of the user-defined region, during the calculation of $U_{\lambda,\rm LL}(\vec{r})$ , or it does not contribute significantly to the RFED of the last intersected cell, during the processing of direct and scattered radiation.

\end{enumerate}

Conversely, rays can be merged when the ray angular density is too high. That is, when: 
\begin{equation}
\Omega_{\rm HP, EM} < \frac{\Omega_{\rm INT}}{\rm N_{\rm rays}^{\rm max}},
\label{equ_omega_low}
\end{equation}
where  N$_{\rm rays}^{\rm max}$ is the user-defined maximum number of rays allowed to cross a cell. 

The previous version of DART-Ray already contained an algorithm for the ray angular density optimization, but it had the problem that many contributions to the RFED of cells already crossed by rays had to be recalculated several times (see NA14 for details).  Instead, in DART-Ray V2 we implemented an optimization strategy for the ray angular density, in which rays can be split and merged along the path they are following, and it avoids repeating the ray-tracing calculations for cells already intersected with a sufficient number of rays. 

\begin{figure*}
%\hspace{-0.5cm}
\centering
\includegraphics[scale=0.4, trim= 4cm 15cm 2cm 10cm ,clip ]{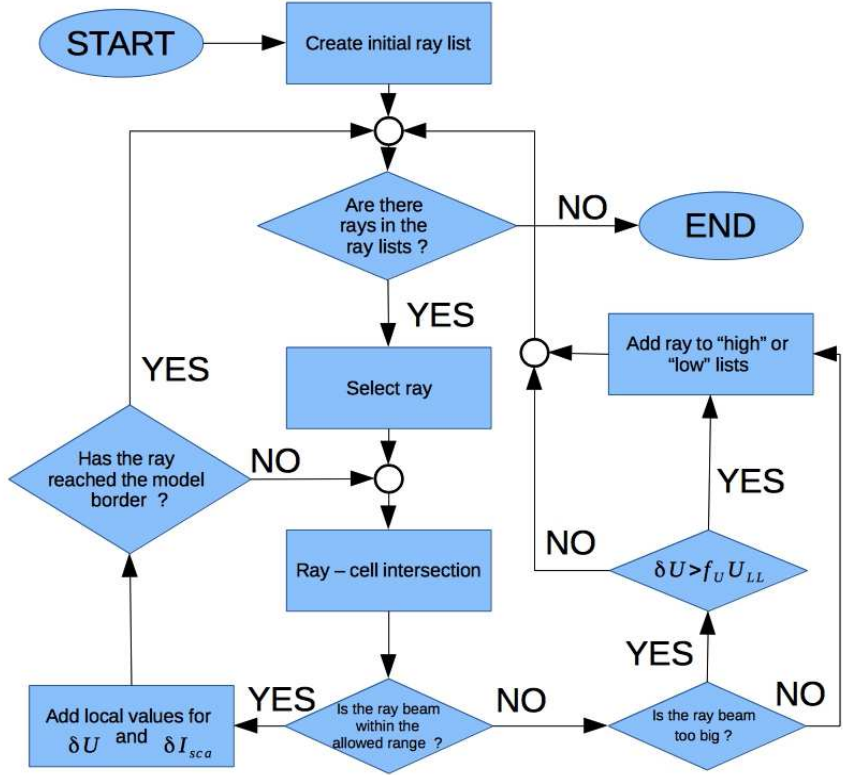}
\caption{Flowchart of the angular density optimization algorithm for the rays originating from a single source. All departing rays are collected in lists (high or low, see text) and their propagation is followed through the RT model. Depending on their beam size and contribution to the RFED, they are split or merged during the `select ray' phase. We note that, during step 1 of the RT algorithm (see Sect. \ref{dartray_algo}), the decision condition $\delta U > f_{\rm U} U_{\rm LL}$ is substituted by $d_{\rm ray} < R_{\rm pre}$ with $d_{\rm ray}$ the distance crossed by the ray and $R_{\rm pre}$ the input maximum distance that can be crossed in this step. } 
\label{fig_flowchart_ang_dens_opt}
\end{figure*}

The method we implemented is similar to the algorithm of \citet{Abel02}, but with several technical differences. This method is described by the flowchart in figure \ref{fig_flowchart_ang_dens_opt}. At the beginning, the code selects a ray and follows its propagation through the RT model. At each ray-cell intersection, it checks whether the ray-beam satisfies any of the conditions expressed in equations \ref{equ_omega_high} or \ref{equ_omega_low}. If not, it adds the ray contributions to the RFED and to the scattering source function. Unless the ray has already reached the model border, the ray propagation continues to the next cell intersection. If the ray beam is found too large after any of these intersections (that is, it satisfies equation \ref{equ_omega_high}), the code checks if the ray still carries a significant contribution to the RFED (see equation \ref{equ_blocking_criteria}). If so, it adds the current ray to the `high' list, the list of rays to be split. Otherwise the ray further propagation is ignored. Instead, if the ray beam is found too small (according to equation \ref{equ_omega_low}), the ray is added to the `low' list, the list of rays that can potentially be merged.    

Once all rays within an HEALPix sector have been processed for the current HEALPix angular resolution, DART-Ray checks whether there are rays in the high ray list. If so, it proceeds with the ray tracing at immediately higher HEALPix resolution. That is, for each ray in the high ray list, four child rays are generated with directions corresponding to the HEALPix directions within the spherical pixel associated with the parent ray. The ray tracing calculation for these child rays starts directly from the same distance $d_{\rm ray}$ from the source which has been already crossed by the parent ray. 

After all rays in the high ray list have been processed, the code looks for rays that can potentially be merged among those in the low ray list. In order to be merged, the directions of four rays in the list should be contained within the same HEALPix spherical pixel at the immediately lower angular resolution, and these four rays should have been blocked after crossing the same grid plane. If so, the code merges them into a single ray with specific intensity equal to the average intensity of the four merged rays. After that, the code starts the propagation of the newly created rays from the average distance crossed by corresponding merged rays. 
The code proceeds with the propagation of all rays from the high and low lists iteratively until there are no more rays in both lists. 

\subsubsection{Wavelength-dependent angular resolution for the scattering source function}
\label{scatt_source_function}

The scattering source function $j_{\rm \lambda,sca}(\vec{r}, \theta,\phi)$ is the computed quantity requiring more memory in the DART-Ray code, since it depends on six independent variables ($\lambda, \vec{r}, \theta, \phi$). By sampling appropriately each dimension, its values can be stored in a big array of size typically in the range $\sim$1-100 Gbytes. Furthermore, the algorithm needs one more array of the same size to store the scattering luminosity to be processed within each scattering iteration, separately from the total scattered luminosity stored in $j_{\rm \lambda,sca}(\vec{r}, \theta,\phi)$. 

As big as it is, storing the scattering source function is still cheaper in terms of memory requirements than solving the radiative transfer equation directly for the specific intensity $I_\lambda(\vec{r}, \theta, \phi)$. This is because $I_\lambda(\vec{r},\theta, \phi)$ can present unpredictably rapid angular variations at each spatial point $\vec{r}$ which are determined by the radiation sources and dust distribution geometry as well as the assumed scattering phase function $\Phi_\lambda(\vec{n}, \vec{n'})$ (dependent on the incoming light direction $\vec{n'}$ and the scattering light direction $\vec{n}$)\footnote{This problem affects all methods which determine $I_\lambda(\vec{r}, \theta, \phi)$ directly, including finite-differencing, discrete ordinates and other ray-tracing methods.}. 
 The latter determines the angular re-distribution of the scattered radiation after each ray-cell intersection. Instead the rapidity of the angular variations for $j_{\rm \lambda,sca}$ is determined only by the shape of $\Phi_\lambda$, typically modelled as a Henyey-Greenstein profile \citep{Henyey41}: 
\begin{equation}
\Phi_\lambda(\vec{n}, \vec{n'}) = \frac{1-g_\lambda^2}{4\pi[1+g_\lambda^2-2g_\lambda\cos\theta]^{3/2}},
\label{equ_HG_function}
\end{equation}
with $\theta$ being the angle between $\vec{n'}$ and $\vec{n}$, and the anisotropy parameter $g_\lambda$  determining the angular width of the $\Phi_\lambda(\vec{n}, \vec{n'})$ profile. For typical interstellar dust models, this profile is quite sharp at UV wavelengths, but it gradually becomes flatter going towards the NIR and then almost completely flat in the FIR. Therefore, the number of angular points needed to sample $j_{\rm \lambda,sca}(\vec{r}, \theta,\phi)$ properly has to be quite high at shorter wavelengths, while relatively few points are sufficient in the FIR where scattering is essentially isotropic. This property of $j_{\rm \lambda,sca}(\vec{r}, \theta,\phi)$ allows a significant reduction in the memory requirement if the storage of $I_\lambda(\vec{r},\theta, \phi)$ is not needed, as in DART-Ray.  

The sampling points for the angular directions of $j_{\rm \lambda,sca}(\vec{r}, \theta,\phi)$ are those of a discretized HEALPix sphere with a total number of pixels equal to $N_{\rm pix} = 12N_{\rm side}^2$ with $N_{\rm side} = 2^{k_{\rm HP}}$ and $k_{\rm HP}$ a positive integer value \citep[see][]{Gorski05}. 
In DART-Ray we implemented the following formula to derive an appropriate $k_{\rm HP}$ for the scattering source function at each wavelength:

\begin{equation}
k_{\rm \lambda, HP} = \frac{1}{2}\log_2 \left( \frac{4\pi}{12\theta_{\rm \lambda, min}^2}\right),
\label{equ_k_hp}
\end{equation}
with $\theta_{\rm \lambda, min}$, the pixel angular size for the required minimum angular resolution, given by:
 
\begin{equation}
\theta_{\rm \lambda, min} = \frac{\rm{FWHM}[\Phi_\lambda - \Phi_\lambda(\pi)]}{n_{\rm FWHM}},
\end{equation}
with $\rm{FWHM}[\Phi_\lambda - \Phi_\lambda(\pi)]$ the Full Width Half Maximum of $\Phi_\lambda$ minus its `background' value at $\theta = \pi$ (in turn depending on $g_\lambda$) and $n_{\rm FWHM}$ the minimum number of pixels within the $\rm{FWHM}$\footnote{Formula \ref{equ_k_hp} can be found by inverting the following equivalence between the approximate pixel solid angle $\theta^2_{\lambda,\rm min}$ required for the minimum angular resolution and the exact pixel solid angle, equal to the ratio of the total solid angle divided by the number of spherical pixels: $$ \theta^2_{\lambda,\rm min} \approx  \frac{4\pi}{\rm N_{\rm pix}}$$.}. 
We found that by choosing $n_{\rm FWHM}=5$ a good accuracy is reached for the benchmark models examined in Sect. \ref{benchmark_comparison}. We note that the values of  $k_{\rm \lambda, HP}$ have to be integers, so the result of formula \ref{equ_k_hp} is approximated to its integer part. A maximum allowed value $k_{\rm \lambda, HP}=2-3$ has to be set to avoid very high memory requirements for very narrow $\Phi_\lambda$ profiles at short UV wavelengths.  

\begin{figure}
\centering
\includegraphics[scale=0.25, angle = 90]{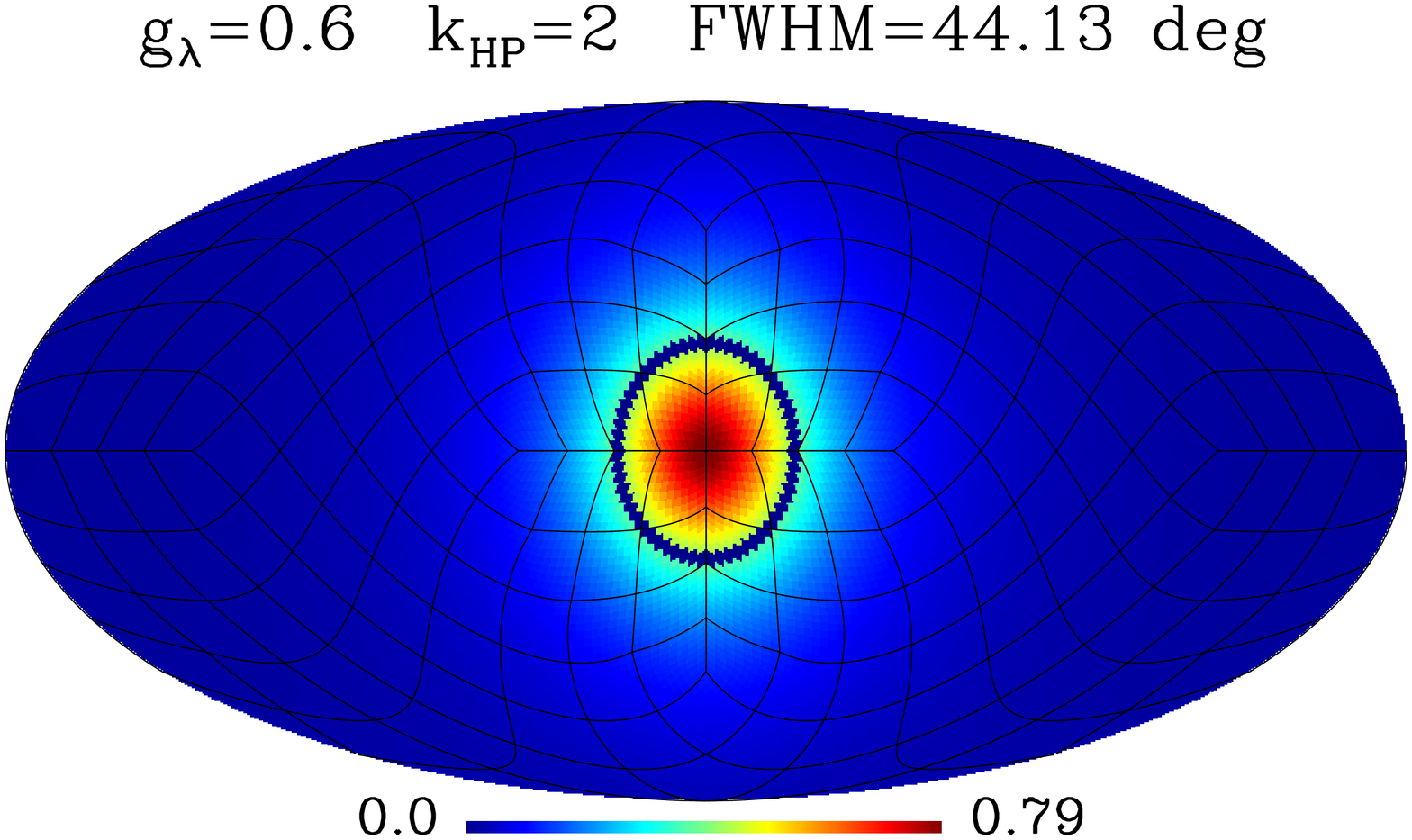}
\includegraphics[scale=0.25, angle = 90]{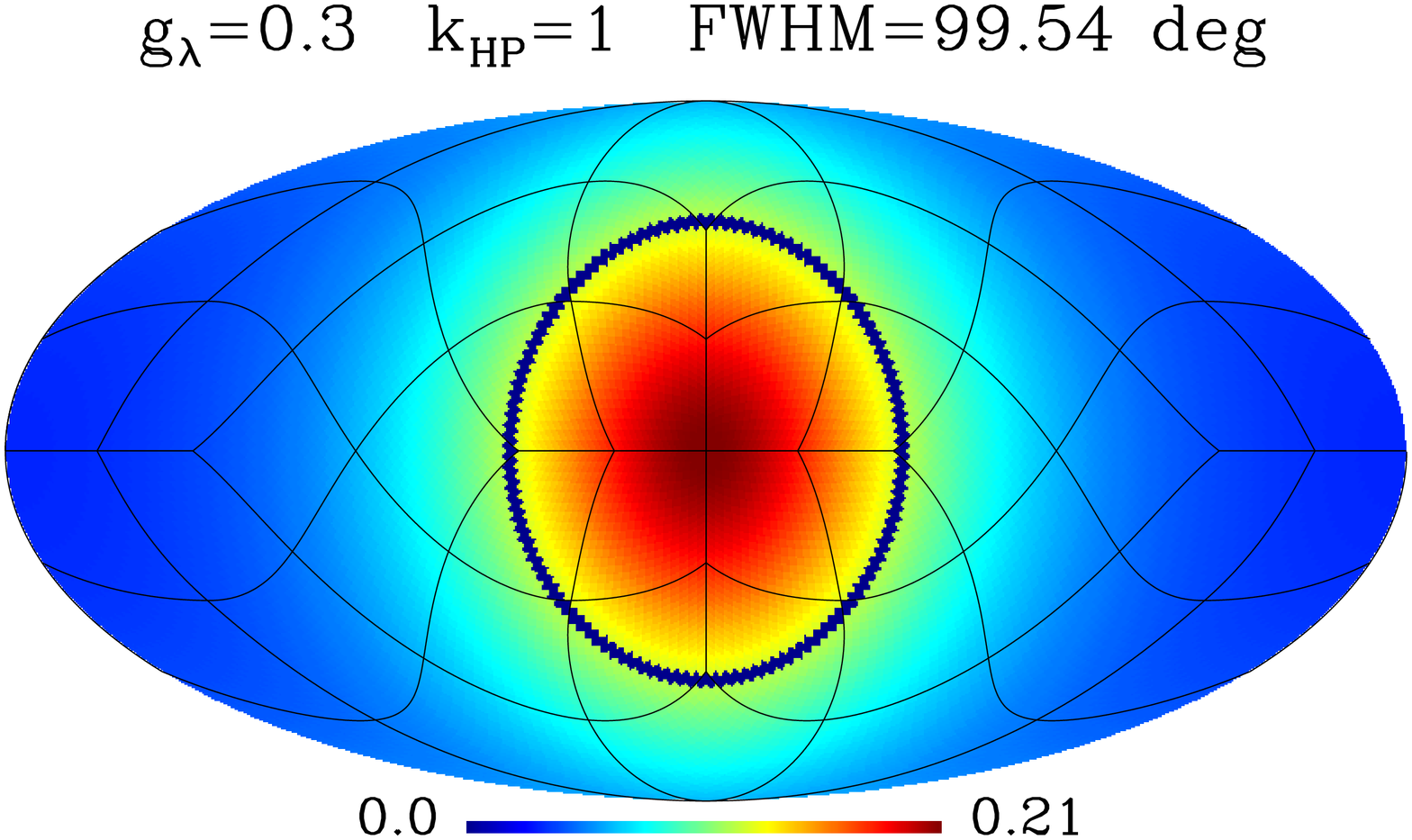}
\includegraphics[scale=0.25, angle = 90]{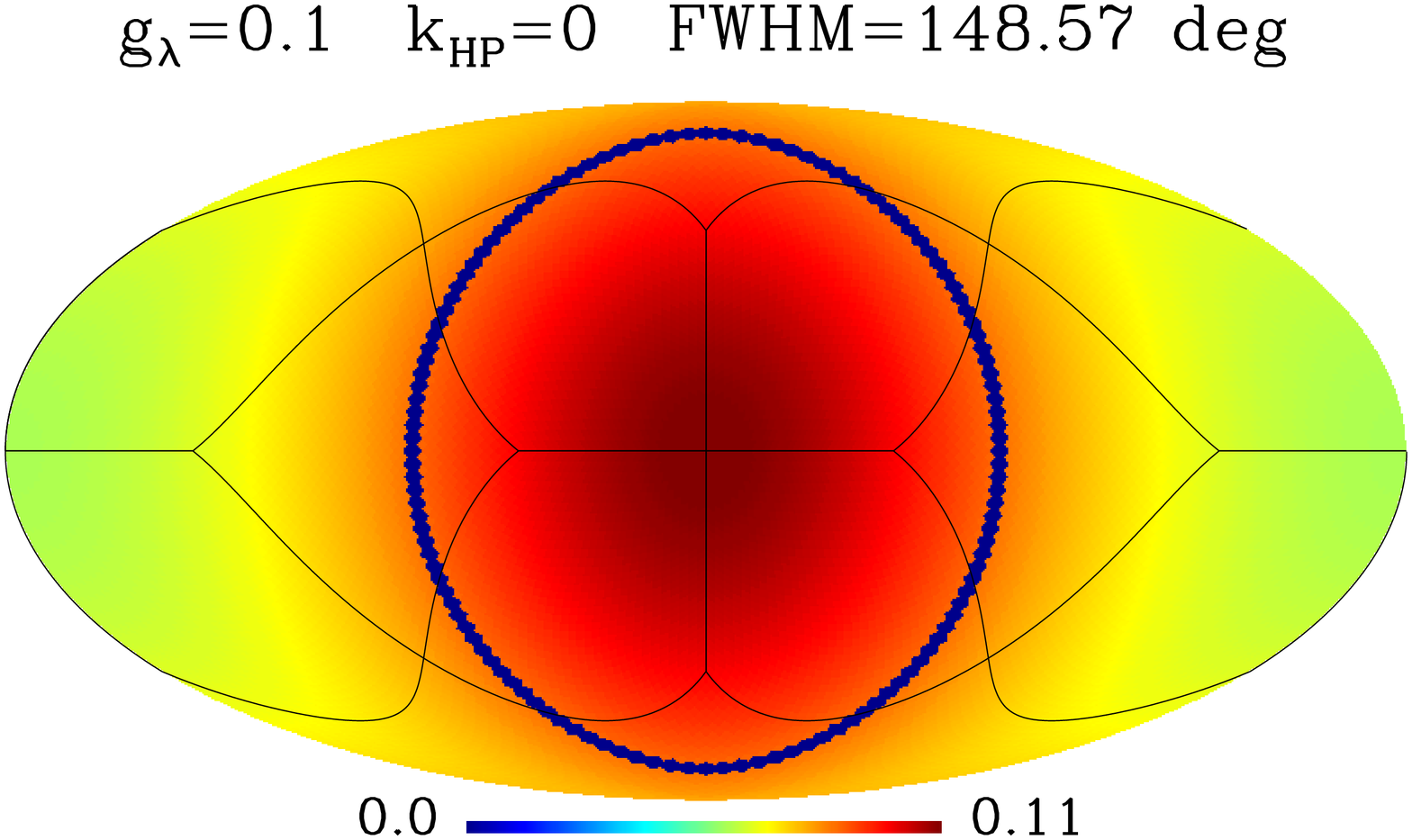}
\caption{Examples of the implemented sampling of the Henyey-Greenstein phase function $\Phi_\lambda$ for different values of the anisotropy parameter $g_\lambda$. The $\Phi_\lambda$ values on a sphere are visualized using Mollweide projection and according to the colour bar below each panel. The contours of the HEALPix pixels corresponding to the values of $k_{\rm \lambda, HP}$ from equation \ref{equ_k_hp} are overplotted as black lines. The dark blue circle shows the size of the $\rm{FWHM} [\Phi_\lambda - \Phi_\lambda(\pi)]$.}
\label{fig_sampling_hg}
\end{figure}

Examples of this sampling can be seen in figure \ref{fig_sampling_hg} for values of $g_\lambda$ in the range 0.1-0.6 (approximately the value range typical of the NIR to UV wavelength range). The figure shows the Henyey-Greenstein functions plotted over the entire sphere using Mollweide projection, together with the contours of the HEALPix pixels for the derived values of $ k_{\rm \lambda, HP}$. Our implementation guarantees that at least several points are sampling the peak of the Henyey-Greenstein profile, which convolves any scattered light contribution added to $j_{\rm \lambda,sca}(\vec{r}, \theta,\phi)$.

The variable angular resolution for the scattering source function allows a considerable reduction in memory. For example, in the TRUST benchmark slab model (see G17) the `BASIC' lambda grid contains 31 wavelengths from the UV until 60 $\mu$m, the range we used for the stellar emission RT. By assuming  $k_{\rm \lambda, HP}=2$ at all wavelengths, and given about 700.000 3D grid points, the memory requirement for  $j_{\rm \lambda,sca}(\vec{r}, \theta,\phi)$ is about 33 Gbytes. By using the variable angular resolution described above, this is reduced to about 12 Gbytes. 

\subsection{Multi-wavelength RT}

Implementing multi-wavelength calculations in a purely ray-tracing code is harder than for MC codes because of the high memory requirement for both the specific intensity $I_\lambda(\vec{r}, \theta, \phi)$ and the scattering source function $j_{\rm \lambda,sca}$. However, this task becomes easier once the memory requirements are reduced when $I_\lambda(\vec{r}, \theta, \phi)$ does not have to be stored in memory and by using a variable angular resolution for $j_{\rm \lambda,sca}$. Because of this latter optimization, DART-Ray V2 can perform multi-wavelength RT calculations without exceeding the RAM memory of modern computer cluster nodes typically of the order of 100 Gbytes. This addition allows the inclusion of dust self-heating (see Sect. \ref{self_heating}) which, being non local in wavelength, cannot be easily handled using a succession of monochromatic RT calculations. Furthermore, many ray-tracing steps are exactly the same for all wavelengths and they are not repeated in multi-wavelength runs. 

Since rays carry multi-wavelength intensities, one has to check that the ray blocking criterion (equation \ref{equ_blocking_criteria}) is fulfilled at all wavelengths. During the ray propagation, this criterion may only be satisfied at some wavelengths. In this case, the code still propagates the ray in the current direction but it does not add the contributions to the RFED and $j_{\rm \lambda, sca}$ at the wavelengths for which the intensity has become negligible. Since updating $j_{\rm \lambda, sca}$ is demanding computationally, this helps to reduce further the calculation time.

\subsection{Dust self-heating}
\label{self_heating}

The previous version of DART-Ray assumed that the dust emission is always optically thin. Thus, it did not consider the absorption of dust emission at other locations (called dust self-heating) as well as the scattering of dust emission. This assumption is not correct for models which are optically thick in the infrared range. This was the main source of disagreement at infrared wavelengths between DART-Ray and the other codes in the TRUST slab benchmark project for the most optically thick models (see G17). In this section, we explain the implementation of the dust self-heating in DART-Ray V2. The result comparison for optically thick models is shown in Sect. \ref{benchmark_comparison}.   

The dust emission spectra at each position is determined by the RFED (or alternatively the average radiation field intensity), the dust density and the dust absorption coefficient \citep[see e.g.][]{Popescu11, Steinacker13}). In DART-Ray this can be calculated assuming either equilibrium between the dust and the radiation field or by deriving the full stochastically heated dust emission spectra (see \citet{Natale15}, \citet{Camps15} and the code user guide).

The dust emission RT calculation is performed after the stellar emission RT is completed. The main difference between the two calculations is that for the former the emission source spectra depend on the RFED. Therefore, while the stellar emission RT run can be performed only once, following the three steps procedure described in Sect. \ref{dartray_algo}, the dust emission RT requires in principle several iterations of that procedure until the dust emission and the infrared RFED both converge. To handle these dust self-heating iterations, we implemented the following procedure: 
\begin{enumerate}
\item the dust emission spectra are calculated taking into account only dust heating from absorbed stellar emission;
\item a first RT calculation for the dust emission is performed following the RT algorithm described in Sect. \ref{dartray_algo};
\item the dust emission spectra are recalculated taking into account the dust heating due to both absorbed stellar emission and the absorbed dust emission;
\item the difference between the dust emission spectra just calculated $j_d(\vec{r})$ and the ones calculated at the end of the previous dust self-heating iteration $j_d^{\rm prev}(\vec{r}) $ is evaluated: 
\begin{equation}
\Delta j_d(\vec{r}) = j_d(\vec{r}) - j_d^{\rm prev}(\vec{r}); 
\end{equation} 
\item another dust radiative transfer calculation is performed during which only the dust emission luminosity stored in $\Delta j_d(\vec{r})$ is processed. The RT algorithm is performed skipping the calculation of the RFED lower limit $U_{\rm \lambda,LL}$, which is set equal to the RFED calculated in the previous iteration. Also, the RFED $U_\lambda$ and $j_{\rm \lambda,sca}$ are initialized with the corresponding values found in the previous iteration; 
\item steps 3,4,5 are repeated until $\Delta j_d(\vec{r})/j_d^{\rm prev}(\vec{r}) < 1\%$ at all positions and wavelengths. 

\end{enumerate}

As one can see, the dust emission RT iterations are performed without processing the same dust emission luminosity more than one time. For moderately optically thick models, $\Delta j_d(\vec{r})$ tends to be very small compared to $j_d(\vec{r})$ already after the first self-heating iteration.  So, the iterations that follow   proceed much faster compared to the first. We tested the validity of this approach in Sect. \ref{benchmark_comparison}.

\subsection{Parallelization}
\label{parallelization}

3D dust radiation transfer is computationally very expensive independently of the algorithm used. Therefore, most of the more advanced dust radiative transfer codes use parallelization to reduce the time needed for the calculations. 

Task parallelization is straightforward to implement because 3D dust radiation transfer is largely an additive problem. For a given RT model, all the quantities to derive, with the exception of the dust emission source function, are equal to the sum of contributions provided by the radiation from the single sources. Therefore, task parallelization is done by distributing the processing of the radiation sources (or photon packages for MC codes) between different CPUs.    

On shared memory machines, as the single nodes of a typical computer cluster, one can easily parallelize the loops over the radiation sources using OpenMP. Unlike MPI, OpenMP allows multiple CPUs to operate on shared arrays. In this way, there is no need for replicating any array or distributing arrays among different processes. Then, to take advantage of multiple nodes simultaneously, a hybrid OpenMP+MPI parallelization scheme is a natural choice, since one can use OpenMP for parallelization within a single node and MPI to handle communication between nodes.  

With multiple nodes it is possible to increase substantially the number of CPUs, and so in principle reduce the total computational time. However, in practice, some overheads that are introduced by the time needed for nodes to communicate and to process the exchanged information. These overheads can become significant when data parallelization among nodes is implemented. 

In DART-Ray, the vast majority of the memory consumption is due to the scattered luminosity source function. For this reason, we did not implement any data parallelization for the other arrays (e.g. the 3D spatial grid coordinates) which are all replicated in all nodes. Instead, we have implemented two data parallelization choices for the scattered luminosity source function: a `communication' mode and a `no-communication' mode. 

In the communication mode, the scattered luminosity source function is distributed among the node memories such that each node contains the scattered source function for different sets of wavelengths. The communication between nodes is needed in two cases: firstly to add the $\delta j_{\rm \lambda,sca}$ contribution at all wavelengths after each ray-cell intersection; secondly, during the scattering iterations, the values at all wavelengths of $j_{\rm \lambda,sca}$ for each source are needed by the same node that has to process it. 

In the first case, in order to minimize data exchange, instead of passing the $\delta j_{\rm \lambda,sca}$ contribution to the scattering source function in each node, only the total scattered luminosities (integrated over all angular directions) and the ray directions are passed to the corresponding node. Then, this information is processed to calculate the angular distribution of scattered luminosity to be added to $j_{\rm \lambda,sca}$ for a certain dusty cell. Furthermore, to minimize communication, large packets of data, containing the scattered luminosity contributions due to many ray-cell intersections, are collected within each node and then exchanged between the nodes (see code documentation for more details). 

\begin{figure}
\includegraphics[scale=0.5]{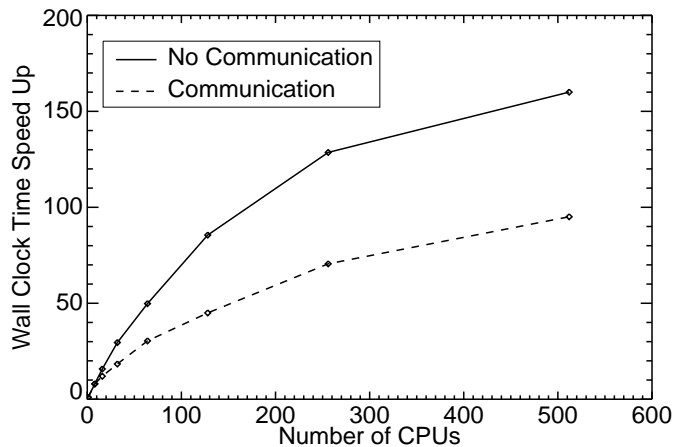}
\caption{Wall clock speed-up with respect to the serial execution for an N-Body and SPH galaxy simulation by using the no communication and communication parallelization mode. For all the parallel runs, we used eight CPUs for each MPI process (that is, eight OpenMP threads per MPI process).  }
\label{fig_dartray_scaling}
\end{figure}

Despite all efforts we put to minimize data exchange and reduce the processing time of the received data, the overheads in the communication mode can still be substantial. Therefore, we also implemented a simpler  no-communication mode where all arrays are replicated in each node, including the scattering source function. In this mode, data communication is performed only at the end of the radiation source loops to sum up the arrays calculated separately by all nodes. An example of the speed-up performances of the communication and no communication parallelization modes is shown in figure \ref{fig_dartray_scaling}. In this figure, we show the wall clock speed-up of the calculation, compared to the serial execution, for the N-body and SPH galaxy model example contained in the DART-Ray current release (see DART-Ray User Guide). As expected, the no communication mode scales much better with the number of CPUs than the communication mode.  However, it is also much more expensive in terms of memory. Nonetheless, with the typical RAM memory of computer cluster nodes of the order of hundreds of Gbytes and thanks to our implementation of the wavelength-dependent angular resolution for the scattering source function, it is possible to use this parallelization mode in the majority of cases. We recommend that users of the code use this mode, unless the memory requirements are so high that data distribution is unavoidable.          

\subsection{Control of the inaccuracy due to ray blocking}
\label{expected_num_accuracy}

There are several factors affecting the numerical accuracy of the RT calculations performed by DART-Ray. Firstly, DART-Ray uses a spatial grid to discretize the distribution of the diffuse stellar emission and dust mass. The RFED distribution as well as the source functions are also evaluated on this same spatial grid. 
However, it is not possible to set the resolution of the spatial grid sufficiently high to attain a pre–defined level of accuracy by the end of the calculation.  While creating the grid, one typically utilizes higher resolution grid elements in regions of higher stellar volume emissivity and dust density. Although reasonable to expect a more rapid variation of the radiation field in those regions, there is no guarantee that the spatial resolution is adequate everywhere on the grid. In the absence of iterative procedures to increase the spatial resolution during the RT calculation, the effect of the grid discretization on the numerical accuracy can only be checked by repeating the calculations at progressively higher spatial resolutions. Similarly, the finite number of angular directions of the rays that are cast from each radiation source, as well as the discretization of the scattering source function, also affect the calculation accuracy. 

Apart from these factors, common to all 3D dust RT codes although in different forms\footnote{Even in MC codes, although photon particles can propagate in any possible direction and be scattered at any location within an RT model, the number of particle directions that can be followed is still finite. This inevitably produces a discretization error which can only be reduced by increasing substantially the number of particles.}, in DART-Ray the numerical accuracy is also affected by the estimate of the extent of the source influence volumes. This is because DART-Ray calculates the RFED contributions from each source only within this volume surrounding the source itself, thus neglecting the contributions outside it. Since this is a core characteristic of DART-Ray, we are interested in quantifying the accuracy error due to the cut off of the rays when they reach the boundary of the estimated source influence volume. We note that this accuracy error is systematic, since it will always produce RFEDs which are underestimated compared to the correct value. 

The ray cut off occurs when the RFED contribution $\delta U_\lambda$ carried by a ray satisfy the criterion \ref{equ_blocking_criteria}. So, once a lower limit to the RFED distribution $U_{\rm \lambda, LL}$ has been estimated, the input--defined parameter $f_U$ is the key factor affecting the numerical accuracy of the calculation. In N14, we stated that $f_U$ should be low enough to preserve energy balance, in the sense that, at the end of the calculation, the total radiation luminosity that has been neglected  because of the ray cut-off should be only a small fraction of the total luminosity of the model. In this case, the effect of cutting the rays is minimal because almost all the radiation luminosity has been followed in its propagation throughout the model. However, since this energy balance can only be checked at the end of the RT run, potentially several attempts had to be made to find the appropriate value for $f_U$. Instead, in the following, we show how the value of the parameter $f_U$ can be set before an RT run such to guarantee the desired level of accuracy. 

We have been able to find a relation between $f_U$ and the accuracy of the derived RFED distribution by making a minor change in the definition of $\delta U_\lambda(\vec{r})$ in formula \ref{equ_blocking_criteria}, compared to NA14. This is now defined as:
\begin{equation}
\delta U_\lambda(\vec{r})=\frac{<I_\lambda> A_{\rm EM} \Omega_{\rm INT} L_{\rm INT}}{V_{\rm INT}~ c},
\end{equation}   
where $<I_\lambda>$ is the average specific intensity of the ray within the ray-cell intersection path, $A_{\rm EM}$ is the area of the emitting cell originating the ray,  $\Omega_{\rm INT}$ is the solid angle subtended by the intersected cell, $ L_{\rm INT}$ is the linear size of the intersected cell,   $V_{\rm INT}$ its volume and $c$ the light speed. This formula differs only slightly from the corresponding formula in N14: the factor $ L_{\rm INT}$ replaces the ray-cell intersection path length, and the factor $\Omega_{\rm INT}$ replaces the ray beam solid angle $\Omega_{\rm HP, EM}$.  So, in the previous version, $\delta U_\lambda(\vec{r})$ in relation \ref{equ_blocking_criteria} was the RFED contribution of the single ray to the intersected cell RFED.  Instead, $\delta U_\lambda(\vec{r})$ now represents approximately the total RFED contribution of the radiation source originating the ray to the intersected cell. In this way, every time the criterion for $\delta U_\lambda(\vec{r})$ is checked, the relevance of the radiation source in determining the local RFED is considered, not just that of the single ray (which can simply have a small intersection path or a small associated $\Omega_{\rm HP, EM}$). 

With this change, an appropriate value for $f_U$ can be derived as follows. When the condition \ref{equ_blocking_criteria} is realized during the ray propagation, one would like to assure that the small contribution $\delta U_\lambda(\vec{r})$ does not sum up with comparable contributions from many other radiation sources which cumulatively provide a non negligible contribution to the intersected cell RFED. In the highly improbable case that all other radiation sources in the RT model provide a RFED contribution as low as $\delta U_\lambda(\vec{r})$, the cumulative contribution $\sum_i \delta U_{\lambda, i}(\vec{r}) $ would be such that:
\begin{equation}
\sum_{i=1}^{N_s} \delta U_{\lambda, i}(\vec{r}) \leq N_s f_U U_{\rm LL}(\vec{r}),
\end{equation}   
where $N_s$ is the total number of radiation sources in the model. By requiring that the RHS of the above inequality is only a small fraction $a_{\rm RT}$ of the final value $U_\lambda(\vec{r})$ for the RFED, we have:
\begin{equation}
N_s f_U U_{\rm \lambda, LL}(\vec{r}) \leq a_{\rm RT} U_\lambda(\vec{r}),
\end{equation}
In the above relation, the factor  $a_{\rm RT}$ represents the desired accuracy of the RT calculation at each position. By assuming conservatively that $U_{\rm \lambda,  LL}(\vec{r})$ is a substantial fraction of $U_\lambda(\vec{x})$, that is $U_{\rm \lambda, LL}(\vec{r})\sim 0.25 U_\lambda(\vec{r})$, we have then a relation between $f_U$ and $a_{\rm RT}$:
 \begin{equation}
 f_U \leq \frac{4a_{\rm RT}}{N_s}.
 \label{equ_f_u}
 \end{equation}
By taking advantage of the above relation, DART-Ray sets the $f_U$ parameter to $f_U = \frac{4a_{\rm RT}}{N_s}$, for a given input-defined accuracy parameter $a_{\rm RT}$. We point out that this input-parameter can be used to control only the inaccuracy due to the blocking of the rays, not the other factors mentioned at the beginning of this section. 

\subsection{Other updates}

We list here other relevant updates to the code. 

\subsubsection{Point sources}

It is now possible to include a set of point sources at arbitrary positions within the 3D grid. This is useful for including unresolved objects, such as stars within a molecular cloud, or an AGN within a galaxy model.

\subsubsection{Use of HDF5 files}

The 3D grid, the output arrays, such as the RFED and the scattering source function, and the surface brightness maps are now be written to files in Hierarchical Data Format\footnote{https://www.hdfgroup.org} (HDF5), although the output can be defined and restricted by the user. HDF5 format offers quicker I/O and smaller file sizes compared to standard ASCII output.

\subsubsection{Internal observer maps}

Surface brightness maps, as seen by an observer within the RT model, can now be produced with DART-Ray. This is useful for creating images and animations for public presentations, and it allows the user to reproduce observations of the Milky Way. The output is in HEALPix format, which is a format used in all-sky surveys, including the recent PLANCK data in the infrared. This feature has been used in \citet{Popescu17} and Natale et al. (2017, in prep) to construct a radiation transfer model of our own Galaxy.

\subsubsection{2D mode}

DART-Ray contains a 2D mode which can be used for axisymmetric models. The calculations are still performed on a 3D cartesian grid but, in this mode, DART-Ray performs the ray-tracing calculations only for the cells located in the first grid octant. Then, taking advantage of the problem symmetries, it derives the RFED and scattering phase function contributions from cells in other octants. This mode is about a factor of 8 times faster than the standard 3D mode.

\section{Comparison with TRUST benchmark solutions at high optical depth}
\label{benchmark_comparison}

DART-Ray has been the only purely ray-tracing code that provided solutions for the first benchmark paper of the TRUST radiation transfer benchmark project (see G17). In that study, several codes have been used to compare the results for a geometry constituted by a dusty slab of uniform density illuminated by a star placed above it. Each code had to provide both total SEDs and images for a set of observer lines-of-sight and a number of wavelengths. Four different models have been considered which varied only for the vertical optical depth of the slab at 1$\mu$m.  In that paper, it is shown that the DART-Ray solutions are in good agreement with all model solutions from the other codes, except for one model, which has the largest optical depth  ($\tau(1\mu m) = 10$, see the TRUST benchmark website for all the comparison plots\footnote{http://ipag.osug.fr/RT13/RTTRUST/BM1.php}). In addition, it was shown that the discrepancy in the dust emission for the most optically thick case was due to the absence of dust self-heating in the old version of DART-Ray. In particular, the lack of scattered dust emission on the images produced large discrepancies between DART-Ray, as well as TRADING, and most of the other codes (see figure 9 in G17). 

\begin{figure*}
\centering
\includegraphics[scale=0.5]{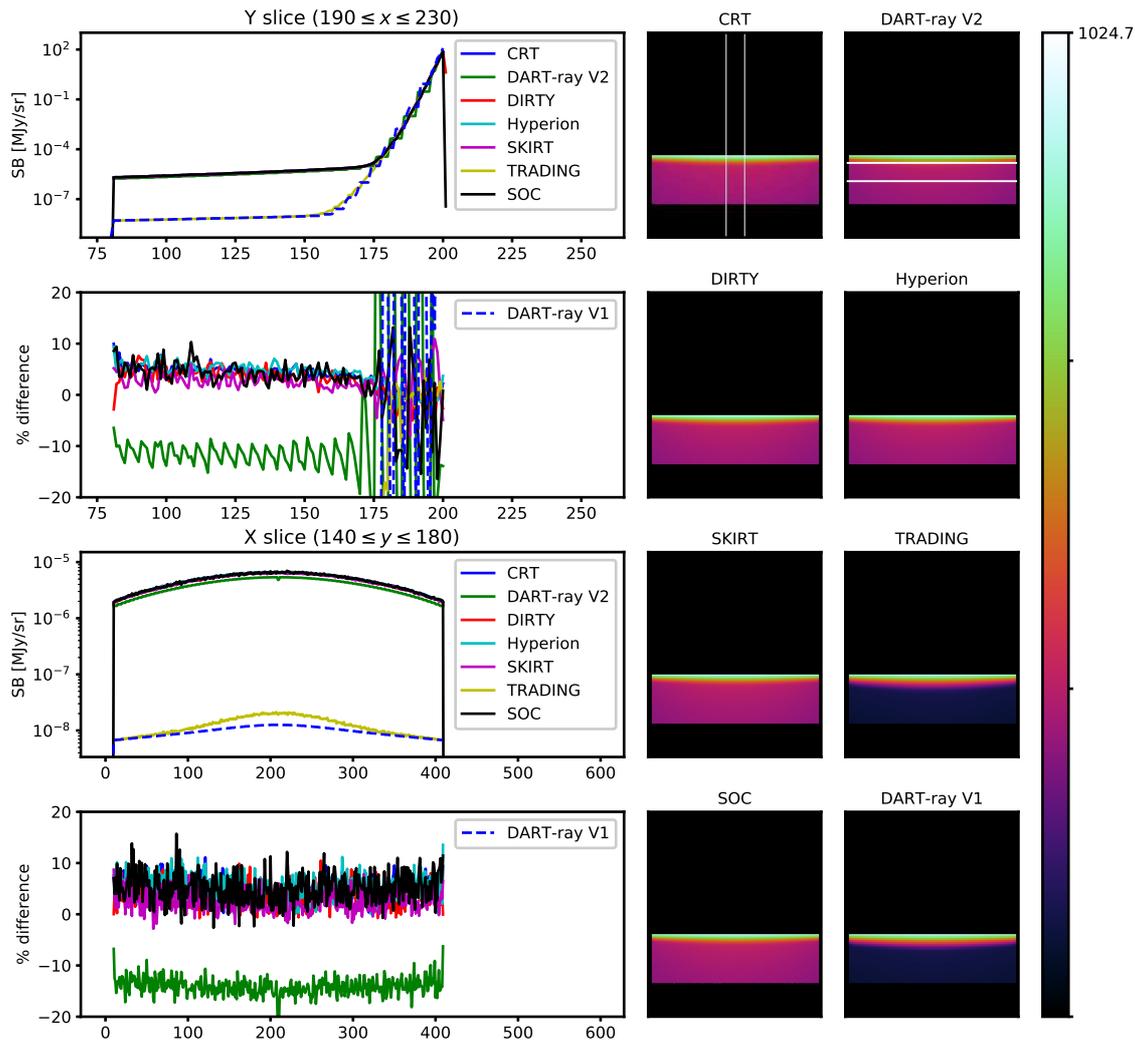}
\caption{Comparison of the $\lambda = 35.11\mu m$ edge-on images of the TRUST slab benchmark for the vertical optical depth $\tau(1\mu m) =10$. The solutions provided by the old and new DART-Ray version are included as well as those of the other codes participating to the project. Units on the images are MJy/sr. The plots on the left show the average surface brightness profiles and the relative differences between the solutions along a vertical and a horizontal strip, whose boundaries are shown within the top two images (CRT and new DART-Ray code). The X-axis of these plots are in units of pixels. The inclusion of dust self-heating in the new version of DART-Ray allows a much closer agreement with the other codes. We note that these solutions are for the `effective grain' case (see G17).}
\label{fig_trust_models}
\end{figure*}

DART-Ray V2 includes dust self-heating as well as dust emission scattering. In order to test that the implementation of these effects is correct, we re-calculated all the TRUST benchmark solutions with the current code. These solutions have now been added to the TRUST website where one can check that they differ at most by $\sim$10\% from the other code solutions in all cases. Here we show only the comparison of the images at $\lambda$= 35.11 $\mu$m, $\tau(1\mu m)$ = 10 and for the edge-on view, which was taken as an example in G17 of the importance of including dust self-heating. This is shown in figure \ref{fig_trust_models} where we included all the other code solutions as well as the old and new DART-Ray solutions. As one can see from the average vertical and horizontal profiles of the surface brightness, DART-Ray V2 produces a MIR image which is much closer to the images of the codes including dust emission scattering. The residual discrepancy is mainly due to the lower spatial resolution of the DART-Ray grid compared to that of the other codes (see G17) together with a contribution due to the ray blocking criterion up to a few percents. We found the same result for all other cases not shown here.  

Apart from images and SEDs, the other main quantity calculated by RT codes is the RFED. Unfortunately, this quantity is more difficult to compare because different codes use different types of grids and resolutions. For this reason, no comparison of the RFED has been made in G17 and the agreement for the dust emission has been taken as an evidence that the RFED have been calculated correctly. We note that the RFED solutions provided by DART-Ray for axisymmetric galaxy models have been compared and found in good agreement with those presented by \citet{Popescu13} (see NA14).

\section{The source influence volume in galaxy models}
\label{source_influence_vol}

The efficiency of the DART-Ray algorithm is based on its criterion to block rays expressed by equation \ref{equ_blocking_criteria}. In the best case scenario, this criterion is satisfied after the rays have crossed only a small part of the model. This would allow the RT calculations to proceed rather quickly. Instead, if the rays have to cross a large fraction of the model before being blocked, the DART-Ray algorithm becomes inefficient. It is therefore interesting to measure the lengths after which the rays are blocked compared to the model size. In this section, we show an analysis of the distribution of these crossed lengths for the Milky Way (MW) galaxy model presented in \citet{Popescu17}. The analytical formulae describing the distribution of stars and dust opacity at each wavelength for this model can be found in that paper.  

\begin{figure*}
\centering
\includegraphics[scale=0.6]{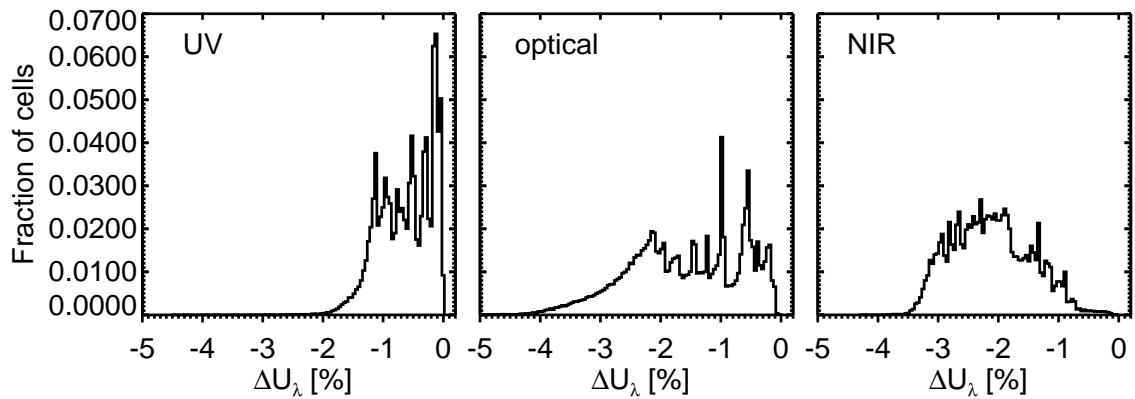}
\caption{Distributions of the relative number of cells as a function of the relative discrepancy for the RFED for the MW model derived by assuming $a_{\rm RT} = 0.005$ and $a_{\rm RT} = 0.05$. The plots show the results for the UV, optical and NIR wavelengths used in the RT calculations. We note that the relative discrepancy is always lower than 5\% as expected. }
\label{plot_hist_delta_ufield}
\end{figure*}

As discussed in Sect. \ref{expected_num_accuracy}, the size of the source influence volume (and thus the lengths at which the rays are blocked) depends on the numerical accuracy that has to be reached in the RT calculation. For the tests presented in this section, we set a maximum numerical inaccuracy to 5\%. This can be achieved by setting $a_{\rm RT} = 0.05$ since the code uses equation \ref{equ_f_u} to set the threshold parameter $f_U$. In order to check that equation \ref{equ_f_u} can be used to set the maximum inaccuracy correctly, we also performed a much more accurate calculation with $a_{\rm RT} = 0.005$ and compared the results for the RFED for a UV (0.150$\mu$m), an optical (0.443 $\mu$m) and a NIR (2.2$\mu$m) wavelength. The distribution of the cells as a function of the relative discrepancy for the RFED for these two different calculations is shown in figure \ref{plot_hist_delta_ufield}. As one can see, the relative discrepancy is never higher than 5\% in absolute values, proving that the accuracy prescription used to set the threshold parameter $f_U$ works correctly. 

\begin{figure*}
\centering
\includegraphics[scale=0.7, angle = 90, trim = {1.5cm, 0, 0, 0}, clip]{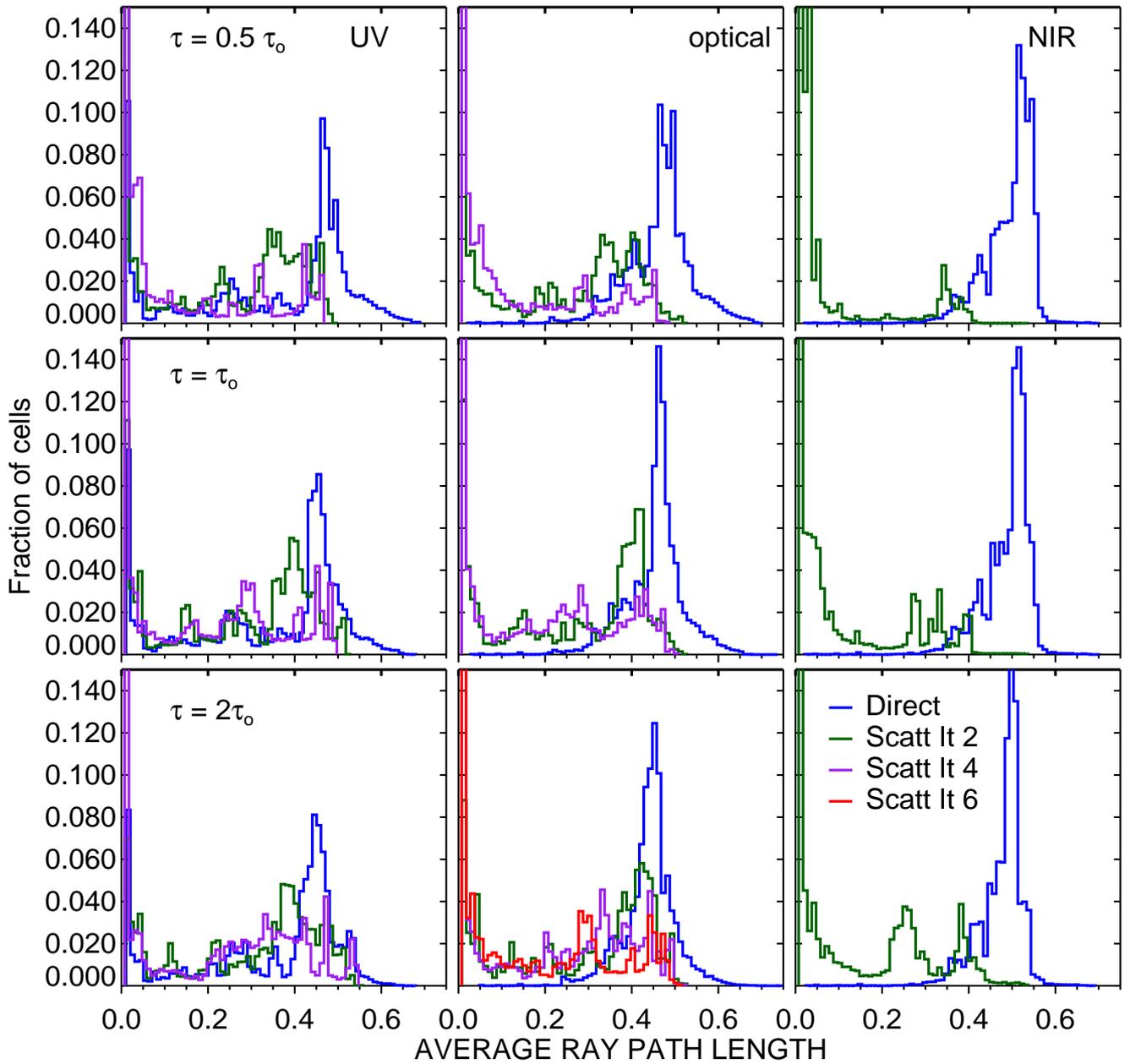}
\caption{Distributions of the relative number of cells as a function of the average ray length crossed by the rays from each cell, normalized to the model linear size, for the direct light processing phase and for some of the scattering iterations. The results are shown for each UV, optical, NIR wavelengths used in the calculations and for the optical depths scaled by the factor 0.5, 1, and 2 compared to the original MW model.}
\label{plot_hist_psel_av}
\end{figure*}

By assuming $a_{\rm RT} = 0.05$, we derived the distribution of the average path crossed by rays departing from each cell for the Milky Way model at the UV, optical and NIR wavelengths mentioned above. We derived this distribution for the direct light processing phase as well as for each scattering iteration. Also, in order to see the effect of varying optical depth on the distributions, we also calculated them for Milky Way models with the dust opacity distribution artificially scaled by the factor 0.5 and 2. All these distributions are shown in figure \ref{plot_hist_psel_av}. We note that the ray path lengths are expressed in units of the model linear size. Also, not all the scattering iteration distributions are shown in order to make the histograms clearer. The median values for all distributions are shown in table \ref{table_influence_volume}. 

From figure \ref{plot_hist_psel_av} and table \ref{table_influence_volume} a few conclusions can be drawn about the sizes of the source influence volumes for each cell and for each calculation phase. Independently of the optical depth, the sizes of the source influence volume are the highest for the direct light processing phase where they can be of the order of half of the model linear size or more. However, the volume sizes decrease with the order of the scattering iterations. In particular, the decrease is rather steeper in the NIR infrared wavelength than that at UV and optical wavelengths. We note that for the NIR wavelength the sizes of the influence volume for the direct light processing are the highest while, at the same time, they shrink rapidly with the order of the scattering iterations. This is because both the optical depth and the albedo of the models at NIR wavelengths are much smaller compared to UV and optical wavelengths. In fact, because of the lower optical depth in the NIR, the ray specific intensity decreases less rapidly during the ray propagation, and the collective contributions to the RFED by many cells at large distances are more important. This makes the source influence volumes larger for the direct light. At the same time, the scattered light has much lower intensity compared to the direct light and does not contribute much to the RFED far away from the dusty cells that originate it. Therefore, the influence volume sizes for the scattered light become small very quickly. 

The effect of increasing the optical depth for the same geometry seems to be different for the direct light and the scattered light iterations. For the direct light, the sizes of the source influence volumes do not change much. Instead, for the same scattered light iteration, the influence volumes seem to be larger with increased opacity. There is no simple explanation for this effect. On one hand increasing the opacity increases the efficiency of scattering light. On the other, it also reduces more rapidly the ray intensity and thus the contribution to the RFED and to the scattered light intensity at large distances. The overall effect seems to be an enlargement of the source influence volumes as well as an increase of the number of scattering iterations required to complete the RT calculation.

\begin{table*}
\centering
\begin{tabular}{|c|c|c|c| c|c|c| c|c|c|}
\hline
 & \multicolumn{3}{|c|}{ 0.5$\tau_0$} & \multicolumn{3}{|c|}{ $\tau_0$} & \multicolumn{3}{|c|}{ 2$\tau_0$} \\
\hline
&  UV & opt & NIR & UV & opt & NIR & UV & opt & NIR \\ 
DIRECT &     0.45 &   0.47 &   0.51 &   0.44 &  0.47 &   0.50 &  0.42 &  0.45 & 0.49   \\ 
SCA IT 1 &   0.38 &   0.38 &  0.24 &   0.39  &  0.39 &   0.27 &  0.39 &  0.40 & 0.30   \\
SCA IT 2 &   0.30 &   0.29 &  0.02 &   0.34  &  0.33 &   0.03 &  0.36 &  0.36 & 0.07   \\
SCA IT 3 &   0.20 &   0.17 &          &   0.28  &  0.26 &           &  0.32 &  0.31 &        \\
SCA IT 4 &   0.04 &   0.05 &          &   0.18  &  0.18  &          &  0.27 &  0.27 &          \\   
SCA IT 5 &           &           &          &            &  0.05 &           &  0.19 &  0.22 &          \\   
SCA IT 6 &           &           &          &            &           &          &          &  0.10 &           \\
SCA IT 7 &           &           &          &            &           &          &          & 0.03  &           \\
\hline
\end{tabular}
\caption{Median values of the average ray crossing length distribution for the Milky Way models with the optical depth scaled by the factors 0.5, 1 and 2. For each model the median values are given for the UV, optical and NIR wavelengths and for the direct light processing phase as well as each scattering iteration. The values are in units of the model linear size.}
\label{table_influence_volume}
\end{table*}

\section{Pro and cons of the DART-Ray code}
\label{Pro_and_cons}

The DART-Ray code is one of the few 3D dust RT codes which do not use the MC method (see \citet{Steinacker13}) and the only one using an algorithm based on estimating the source influence volume extents. The originality and the relative novelty of this code are accompanied by several advantages and disadvantages: \\

\textbf{Advantages}
\begin{itemize}
\item no MC noise; 
\item RT calculation very efficient for higher order of the scattered light; 
\item it calculates the radiation field energy density accurately everywhere, even when its knowledge is not required to produce images;  
\item alternative method that can be used to validate further scientific results obtained by MC codes; 
\item allows to calculate images at arbitrary observer positions without repeating the entire RT calculation;
\item flexibility to change input geometry, dust model, stellar emission library; 
\item easy to import N-body and SPH simulations in tipsy format.
\end{itemize}

\textbf{Disadvantages}
\begin{itemize}
\item high memory requirements;
\item lack of subgrid resolution (exploited by MC codes); 
\item typically longer calculation times compared to those of MC codes; 
\item direct light calculation rather inefficient when the source influence volumes are close to the entire RT model;
\item only Cartesian adaptive grid implemented.
\end{itemize}

\section{Possible further improvements}
\label{sec_improvements}

DART-Ray V2 is a major improvement compared to the code presented in NA14. Apart from the new capabilities, the code has now a solid structure and documentation that makes further development possible. The main barrier to overcome in DART-Ray is the reduction of the calculation time for models in which the sources have influence volume sizes of the order of the entire model size. For example, this typically happens for galaxy models in the infrared range, where the galaxy is more transparent and sources cumulatively contribute to the RFED at large distances. However, in the same models, the sources of scattered light tend to have rather small influence volumes and, therefore, the processing of scattered light proceeds much faster. A more efficient algorithm could be built which processes the direct light in a more efficient way than the adopted source-to-cell approach, while leaving the algorithm as it is for the scattered light processing.

\section*{Acknowledgements}
G.N. and C.C.P. would like to acknowledge support from the Leverhulme Trust research project grant RPG-2013-418.  VPD is supported by STFC Consolidated grant \#ST/M000877/1. G.N. thanks Dimitris Stamatellos for useful comments that helped improve the paper and Karl Gordon for uploading the solutions presented in this paper to the TRUST project website. We thank the anonymous referee for insightful comments on the code algorithm.

\bibliographystyle{aa} % style aa.bst
\bibliography{rtcode_paper3} % your references Yourfile.bib

\end{document}